*The Guilty (Silicon) Mind: Blameworthiness and Liability in Human-Machine Teaming*

Dr Brendan Walker-Munro and Dr Zena Assaad*

As human science pushes the boundaries towards the development of artificial intelligence (AI), the sweep of progress has caused scholars and policymakers alike to question the legality of applying or utilising AI in various human endeavours. For example, debate has raged in international scholarship about the legitimacy of applying AI to weapon systems to form lethal autonomous weapon systems (LAWS). Yet the argument holds true even when AI is applied to a military autonomous system that is not weaponised: how does one hold a machine accountable for a crime? What about a tort? Can an artificial agent understand the moral and ethical content of its instructions? These are thorny questions, and in many cases these questions have been answered in the negative, as artificial entities lack any contingent moral agency. So what if the AI is not alone, but linked with or overseen by a human being, with their own moral and ethical understandings and obligations? Who is responsible for any malfeasance that may be committed? Does the human bear the legal risks of unethical or immoral decisions by an AI? These are some of the questions this manuscript seeks to engage with.

.

## Introduction

Automation has been a key result of mankind's technological development over the last two centuries. Rather than a reliance on manual labour, as a society we have developed mechanised tools which replace our efforts with streamlined and optimised acts, preferably themselves undertaken by machines. Even in the most sensitive and value-driven theatre of human endeavour – that of decision making – the march of progress has not slowed, such that we now have computer programs capable of making decisions on everything from restaurant orders and hotel bookings to delivery of healthcare and social welfare programs.[1]

---

* <AUTHOR DETAILS>. The research for this paper received funding from the Australian Government through the Defence Cooperative Research Centre for Trusted Autonomous Systems.
[1] Igor Bikeev, Pavel Kabanov, Ildar Begishev, Zarina Khisamova, 'Criminological Risks and Legal Aspects of Artificial Intelligence Implementation' (*Proceedings of the International Conference on Artificial Intelligence, Information Processing and Cloud Computing*, Sanya, December 2019).

Yet that automation is not without its controversy. Discussion has raged in the international community regarding the legitimacy of merging the "hard" processing capabilities of a computer with the "soft" processing abilities of a human.[2] Whilst the reality of such a concept might previously have been restricted to the pages of popular fiction,[3] this is no longer the case. Human-machine interfaces – where a system operates to modulate a human's sensory connection with a machine – are already being used in contemporary applications such as piloting drones and other autonomous and semi-autonomous platforms.[4] Scholars are now examining the next step of this inclusion of machines in the human realm of decision-making with an increased research interest in "human-machine teaming" (HMT).[5]

Conceptually, HMT will take human-machine collaboration to another level by more closely fusing the processing capabilities of machines and humans – yet despite the research interest, the literature still lacks a cohesive framework which adequately reflects the legal responsibility for HMT. Imagine a car assembly, where human workers and robotic workers complete their tasks side-by-side, assembling the components of a vehicle as part of a smoothly operating team. However, both the humans and machine are also given a particular values framework imposed by the factory owner: vehicles must be completed to a certain standard, within a certain time. What happens when the machines realise that their human counterparts are the ones that are slowing down the process, making mistakes, costing time and resources? A human worker might seek to disobey the restrictions imposed on him or her by the factory owner – to strike, or maybe just going at their own pace and risking dismissal. Robots have no such flexibility in their programming: what happens if they decide, for coldly logical reasons, it would be more efficient to kill off their human co-workers?

---

[2] Linda Skitka, Kathleen L. Mosier, Mark Burdick, 'Does automation bias decision-making?' (1999) 51(5) *International Journal of Human-Computer Studies*, 991; Ericka Rovira, Kathleen McGarry, Raja Parasuraman, 'Effects of imperfect automation on decision making in a simulated command and control task' (2007) 49(1) *Human Factors,* 76; Gustav Markkula, Richard Romano, Ruth Madigan, Charles W. Fox, Oscar T. Giles, Natasha Merat, 'Models of human decision-making as tools for estimating and optimizing impacts of vehicle automation' (2018) 2672(37) *Transportation Research Record*, 153; Monika Zalnieriute, Lyria Bennett Moses, George Williams, 'The rule of law and automation of government decision-making' (2019) 82(3) The Modern Law Review, 425.
[3] Alan Turing, 'Computing Machinery and Intelligence' (1950) 59 *Mind* 236, 433-460.
[4] Jennifer Riley, Laura D. Strater, Sheryl L. Chappell, Erik S. Connors, Mica Endsley, 'Situation Awareness in Human-Robot Interaction: Challenges and User Interface Requirements', in Michael Barnes and Florian Jentsch (Eds.), *Human-Robot Interactions in Future Military Operations* (CRC Press, Boca Raton), 180.
[5] The term "human-machine team" and "human-machine teaming" are functionally the same for present purposes, and can be used interchangeably throughout this paper.

This might sound like the plot to a particularly ridiculous Hollywood blockbuster – yet some semblance of these facts can be found in reality. Kenji Urada is widely recognised as the first human to "die by robot". In 1981, Urada was performing maintenance on an automated hydraulic arm which, despite written safety protocol, was still powered on. The system misinterpreted Urada's actions as an attempt to damage the arm, which reacted by knocking Urada into an adjacent machine. Urada was crushed and died instantly.[6] A similarly horrifying (though less serious) incident occurred in 2022 when a 7-year-old chess player had his finger broken by a robotic opponent.[7] In both cases, blame was laid squarely on the human for violating safety protocol, and otherwise no charges were laid and no justice was served.

Nowhere should this development be more concerning than in the field of the military and armed forces given the rapid development of research into the 'deployment of AI-infused systems (e.g. drone swarming, command and control decision-making support systems and a broader range of autonomous weapon systems)'.[8] Whilst the idea of HMT presents obvious benefits to military operations, the controversy arises by inflaming existing risks or generating new challenges. Of relevance to military commanders and systems designers – also the thesis of this article – should be a conceptual question about the attribution of responsibility for unlawful actions committed within HMT operations: do the actions give rise to civil liability (where the remedy is usually compensation or some remedial order of the court) or criminal liability (where the remedy is usually imprisonment for natural persons, both as a form of punishment and to protect innocent members of society).

We therefore set out in this article to advance the proposition that, for HMT, the specifics of the dynamic interactions between the human and machine elements will dictate how liability will be attributed. For the context of this paper, HMT is defined as a bi-directional combination of human and machine capabilities which work together with a dynamic directedness towards an aligned goal.[9] We intend to approach the problem in the following way. Part I will involve an exploration of the issues of HMT operations. This Part will identify that the bi-directionality of communication between the human and machine elements serves to blur the perceptions and observations of both parts, and may have legal and regulatory ramifications. Part II will then introduce some key terms in the context of both civil and criminal law around the establishment of liability, with reference to the idea of blameworthiness. In Part III, we explore how a specific mechanism of approaching blameworthiness and liability might be conducted in the future of HMT.

---

[6] Yueh-Hsuan Weng, Chien-Hsun Chen, Chuen-Tsai Sun, Toward the Human-Robot Co-Existence Society: On Safety Intelligence for Next Generation Robots (2009) 1 *International Journal for the Society of Robotics*, 273.

[7] Jon Henley, "Chess robot grabs and breaks finger of seven-year-old opponent", *The Guardian* (online, 24 July 2022) <https://amp-theguardian-com.cdn.ampproject.org/c/s/amp.theguardian.com/sport/2022/jul/24/chess-robot-grabs-and-breaks-finger-of-seven-year-old-opponent-moscow>.

[8] James Johnson, 'The AI-cyber nexus: implications for military escalation, deterrence and strategic stability' (2019) *Journal of Cyber-Policy*, https://doi.org/10.1080/23738871.2019.1701693, 1.

This Article will also specifically focus on HMT in a military context. There are three reasons for such a focus. The first is that HMT is a significant component of the technological research for many Western military forces including the US, UK and Australia,[10] but also of other nations such as China.[11] Secondly, like their comparative cousins in the form of autonomous weapon systems, the application of AI to military decision making in HMT is already recognised as a challenge to the rules-based order of international and comparative domestic law.[12] And thirdly, the military are often a testbed for emerging technologies, with armed forces standing as the entity which commonly responds to the legal and regulatory challenges that arise from their implementation.[13]

---

[9] Zena Assaad, work in progress.

[10] Ministry of Defence, *Human Machine Teaming* (Joint Concept Note 1/18, May 2018) <https://www.gov.uk/government/publications/human-machine-teaming-jcn-118>; Chad C. Tossell, Boyoung Kim, Bianca Donadio, Ewart de Visser, 'Appropriately Representing Military Tasks for Human-Machine Teaming Research', in Constantine Stephanidis, Jessie Y. C. Chen, Gino Fragomeni (Eds.), *HCI International 2020 – Late Breaking Papers: Virtual and Augmented Reality* (Springer, 2020), 245-265; Alex Neads, David J. Galbreath, Theo Farell, *From Tools to Teammates: Human Machine Teaming and the Future of Command and Control in the Australian Army* (Australian Army Occasional Paper No. 7, 20 September 2021).

[11] Department of Defense, *Military and Security Developments Involving the People's Republic of China* (2021), 146-148, <https://media.defense.gov/2021/Nov/03/2002885874/-1/-1/0/2021-CMPR-FINAL.PDF>.

[12] Aiden Warren, Alek Hillas, 'Lethal Autonomous Weapons Systems: Adapting to the Future Unmanned Warfare and Unaccountable Robots' (2017) 12(1) *Yale Journal of International Affairs*, 71; Aiden Warren, Alek Hillas, 'Friend or frenemy? The role of trust in human-machine teaming and lethal autonomous weapons systems' (2020) 31(4) *Small Wars & Insurgencies*, 822.

[13] See for example how drone regulation has emerged in military contexts: Ferran Giones, Alexander Brem, 'From toys to tools: The co-evolution of technological and entrepreneurial developments in the drone industry' (2017) 60(6) *Business Horizons*, 875; Matthieu J. Guitton, 'Fighting the locusts: implementing military countermeasures against drones and drone swarms' (2021) 4(1) *Scandinavian Journal of Military Studies*, 1.

**Part I – Definitional issues of human-machine teaming**

One of the most significant challenges facing the academic and industrial community is the lack of a shared definition on exactly what comprises a HMT. Definitions are vitally important for legal and regulatory purposes, not just as academic or theoretical constructs. The blurring of responsibility between the human and machine elements in a HMT – indeed, the very concept of identifying where a human ends and a machine begins – has the capacity to present significant challenges to the legal and regulatory framework for future HMT operations. If a legal principle cannot apply to the emergence of HMT operations, or applies weakly or ambiguously, the danger of an unregulated system is plainly apparent. Even absent the possibility that HMT (especially military HMT) might be operating without a proper form of legal control or oversight, the absence of a proper regulatory system has the capacity to diminish public trust in the operations of the armed forces which deploy such systems. Worse, the deliberate *unregulated* deployment of such systems may in fact expose those same armed forces to liability themselves.[14]

One such example defines a HMT as 'a purposeful combination of human and cyber-physical elements that collaboratively pursue goals that are unachievable by either individually'.[15] Some broader literature of HMT have similarities in their proposed definitions, with many expressing notions of sharing authority to pursue common goals.[16] Such a definition clearly articulates the connection and bi-directionality between the human (natural) and the machine (artificial), yet articulates these by reference to a frame in which goals <u>cannot</u> be achieved by one or the other in isolation. Applying such a definition to the simple act of driving a vehicle highlights the definitional issues – clearly, both humans and machines can operate, steer and control a vehicle without necessary recourse to the other[17].

---

[14] Consider for example the application of Article 36 of Additional Protocol I of the Geneva Convention: Damian P. Copeland, 'Legal Review of New Technology Weapons', in Hitoshi Nasu, Robert McLaughlin (Eds.), *New Technologies and the Law of Armed Conflict* (Springer, 2014) 43-55.

[15] Azad M. Madni, Carla C. Madni. 'Architectural framework for exploring adaptive human-machine teaming options in simulated dynamic environments' (2018) 6 *Systems* 4, 49.

[16] Joseph B. Lyons, Katia Sycara, Michael Lewis, August Capiola, 'Human–Autonomy Teaming: Definitions, Debates, and Directions' (2021) 12 *Frontiers in Psychology*, 1932, DOI 10.3389/fpsyg.2021.589585.

[17] J. Levinson et al., 'Towards fully autonomous driving: Systems and algorithms' (2011) *Proceedings of the IEEE Intelligent Vehicles Symposium (IV)*, pp. 163-168, doi: 10.1109/IVS.2011.5940562.

Another definition of HMT might be of more utility: 'the dynamic arrangement of humans and cyber-physical elements into a team structure that capitalizes on the respective strengths of each while circumventing their respective limitations in pursuit of shared goals'.[18] Is it the existence of collaboration then, of movement towards a shared goal, which hallmarks human-machine teaming? Yet again, the difficulty in the detail surfaces when applied to a contextual application. Imagine a drone equipped with missiles, deployed in a foreign State but monitored in its home State by a human operator. Both the drone and the operator have a shared goal – the identification, pursuit, and engagement of the State's legitimate military targets – but the nature of the relationship is perhaps better characterised as supervision than 'circumventing their respective limitations'. Obviously, the drone is performing a function in replacement of the human operator and takes the risk in doing so, but the operator still is the one who carries the risk associated with commencing or prosecuting any attack.

How then do the various world militaries approach this definitional issue? The Australian Army broadly defines HMT as the 'incorporation of autonomous or robotic systems within military teams to achieve tactical outputs that neither machines nor people could deliver independently',[19] whilst the United Kingdom's joint concept note on HMT defines the 'effective integration of humans, artificial intelligence (AI) and robotics into warfighting systems'.[20] The US Department of Defense does not strictly define HMT, instead referring to it more obliquely via terminology buried in the program definitions. Take for example the Next-Generation Nonsurgical Neurotechnology (N3) project, which:

> …aims to develop high-performance, bi-directional brain-machine interfaces for able-bodied service members. Such interfaces would be enabling technology for diverse national security applications such as control of unmanned aerial vehicles and active cyber defense systems or teaming with computer systems to successfully multitask during complex military missions.[21]

What is common about these military definitions is an incorporation of, or integration between, human and machine components, to achieve outcomes for the armed forces in combat and peacetime.

---

[18] Lyons et al (n 16).

[19] Neads, Galbreath and Farell (n 10).

[20] Ministry of Defence (n 10), 39.

[21] DARPA, *Our Research: N3* (website, 2022) <https://www.darpa.mil/program/next-generation-nonsurgical-neurotechnology>.

Yet these similarities in the military context also betray the same difficulties in the execution of HMT, which is to explain why HMT are an ideal to which military forces aspire. For that purpose, the Australian Army definition contains perhaps the most succinct policy purpose of HMT; that is, to '…achieve tactical outputs that neither machines nor people could deliver independently',[22] a concept directly reflecting the ideal in the literature that HMT ought to circumvent the respective limitations of human and machine.[23]

Similar lack of consistency affects definitions in other research spheres. For example, the report published by the UN Institute for Disarmament Research (UNIDIR) does not explicitly define HMT. Instead, it focuses on defining aspects of the spectrum of HMT operations from 'coactive design' – in which the machine and human operate in tandem but separately achieve their assigned goals – to 'immersion' – in which the machine and human operate in 'virtual worlds that are simulated [and] dynamic'.[24] NATO approaches to HMT also focus not on the definition of the term, but on the supposed benefits to military decision-making, noting that teaming is the ultimate expression of collaboration, trustworthiness and adaptation between the human and machine components.[25]

Yet a critical thread can be observed across both the military and non-military works seeking to define HMT. Contemporary military capabilities already involve collaboration between humans and machines – whether the machine is a sensor, interface, weapon or system –capable of communicating in a shared language, to achieve or move towards some shared goal. The sharing of these capabilities between human and machine is necessary for achieving outputs that neither entity could complete independently. However, the critical thread observed in these definitions (and the one that lies at the core of this article) is the bi-directionality of that communication. A machine may communicate with its human capability by displaying sensor information, or the projected results of a weapon detonation,[26] whilst a human may provide commands to select, track or engage targets presented.[27]

---

[22] Neads, Galbreath and Farell (n 10).
[23] Madni and Madni (n 15); Lyons et al. (n 16).
[24] Ioana Puscas, Human-Machine Interfaces in Autonomous Weapon Systems: Considerations for Human Control (UNIDIR, 2022) 15-16.
[25] Karel van den Bosch and Adelbert Bronkhorst, *Human-AI Cooperation to Benefit Military Decision Making* (NATO Report STO-MP-IST-160), S3-1-8.

This bi-directionality of communication also presents a unique challenge to attributing responsibility to which agent in the team made a particular decision. Autonomous weapons and military robotics have long been suggested to suffer from a "responsibility gap"[28] – the idea that mankind cedes control to machines when they are invested with the capability to learn and evolve – which Matthias postulated would lead to 'injustice of holding men responsible for actions of machines over which they could not have sufficient control'.[29] Ryan writes that 'military organizations must…examine whether it is desirable to have robots able to kill humans based on automated processes and without a human in the decision cycle'.[30] Yet the idea of a conjoined or collaborative HMT sidesteps Ryan's understanding of the issue: the existence of a human "in the loop" of decision-making is no safeguard against failure.

Consider the following scenarios involving hypothetical military HMTs, but which have been based on existing automated or autonomous technologies:

- The pilot of an attack aircraft, assisted by uncrewed sensor drones,[31] attack a convoy based on the drones' assessment of those vehicles as being legitimate military targets. Following an investigation, it is revealed that the convoy contained fleeing refugees and the sensor data was incorrectly interpreted by the drones.

---

[26] See for example the Athena AI which can differentiate between objects protected under international humanitarian law and legitimate targets: Jonathan Bradley, *Athena AI helps soldiers on the battlefield identify protected targets* (website, Create Digital, 26 April 2021) < https://createdigital.org.au/athena-ai-helps-soldiers-identify-protected-targets/>.

[27] Vasja Badalič, 'Automating the Target Selection Process: Humans, Semiautonomous Weapons Systems, and the Assault on International Humanitarian Law', in Aleš Završnik and Vasja Badalič (Eds.) *Automating Crime Prevention, Surveillance, and Military Operations* (Springer, 2021), 223–242.

[28] Thomas Hellström, 'On the moral responsibility of military robots' (2013) 15(2) *Ethics and Information Technology* 99; Merel Noorman, Deborah G. Johnson, 'Negotiating autonomy and responsibility in military robots' (2014) 16(1) *Ethics and Information Technology* 51; Lambèr Royakkers, Peter Olsthoorn, 'Lethal Military Robots: Who Is Responsible When Things Go Wrong?', in Mehdi Khosrow-Pour (Ed.) *Unmanned Aerial Vehicles: Breakthroughs in Research and Practice* (IGI Global, 2019) 394-411; Isaac Taylor, 'Who Is Responsible for Killer Robots? Autonomous Weapons, Group Agency, and the Military-Industrial Complex' (2021) 38(2) *Journal of Applied Philosophy* 320; Bernd W.Wirtz, Jan C. Weyerer, Carolin Geyer, 'Artificial intelligence and the public sector—applications and challenges' (2019) 42(7) *International Journal of Public Administration* 596.

[29] Andreas Matthias, 'The Responsibility Gap: Ascribing Responsibility for the Actions of Learning Automata' (2004) 6(3) *Ethics and Information Technology* 175, 183.

[30] Mick Ryan, *Human-Machine Teaming for Future Ground Forces* (Report, Center for Strategic and Budgetary Assessments, 2018) 36.

[31] Based on the Loyal Wingman project developed by the Royal Australian Air Force: RAAF, 'Loyal Wingman', *Future Air and Space Capability* (website, 2020) <https://www.airforce.gov.au/our-mission/loyal-wingman>.

- The captain of a Naval destroyer is linked to the automated defences of their ship. Radar detects an aircraft approaching and assesses its behaviour as benign, yet the captain believes the aircraft is adopting an attack profile and opens fire. The aircraft was in fact an allied fighter in an adjacent battlegroup[32];
- A platoon of soldiers is conducting a patrol in a foreign country, assisted by an armed robotic companion that is teamed with one of the platoon soldiers.[33] Unbeknownst to the platoon, the software underpinning the robot has been hacked by enemy forces and suddenly presents false threat warnings. The teamed soldier opens fire, killing one of his platoon members.

As shown above, each of these scenarios highlights a specific concern with the attribution of responsibility for a military HMT. In the first scenario, there was no malicious or adverse action by any person, merely the occurrence of what might be called 'human error' – yet it was an error that resulted in the preventable deaths of civilians.[34] In the second scenario, the naval captain imparted his human bias (contradicting the automated assessment of the aircraft's behaviour) into the decision-making cycle, arguably undermining the purpose of a HMT in the first place. And in the last scenario, the addition of a cyber-physical element into human warfighting opens new avenues for misdirection and attack by enemy forces.

In all three scenarios, the issue of bi-directionality is front and centre at the difficulty of attributing responsibility. The pilot in the first scenario may well have been able to avert disaster had he or she not relied on the drones' sensor information and been able to visually observe the target, something required by pilots in previous conflicts.[35] In the second scenario, the communication with the human is what has hampered the machine in (correctly) identifying that the aircraft was not a threat. Inversely, the third scenario demonstrates that the bi-directionality of communication in HMT introduces a vulnerability which can be exploited by adversarial forces. So how will this bi-directionality affect the legal treatment of HMT?

---

[32] Adapted from the downing of a US Intruder aircraft by a Japanese destroyer in 1996: Thomas Newdick, 'The Last Time A Japanese Warship Shot Down A U.S. Navy Plane Was Actually Not So Long Ago', *The Drive* (website, 5 June 2021) <https://www.thedrive.com/the-war-zone/40937/the-last-time-a-japanese-warship-shot-down-a-u-s-navy-plane-was-actually-not-so-long-ago>.

[33] Based in part on the arming of a Boston Dynamics "dog" robot: Joshua Rhett Miller, 'Robot dog equipped with submachine gun is "dystopian" nightmare fodder', *New York Post* (website, 21 July 2022) <https://nypost.com/2022/07/21/robot-dog-with-submachine-gun-is-dystopian-nightmare-fodder/>.

[34] For example, consider the 'human error' that led to the US strike on a Medecins Sans Frontiers hospital in Kunduz, Afghanistan in 2015: John F. Campbell, *Investigation Report of the Airstrike on the Medecins Sans Frontieres / Doctors Without Borders Trauma Center in Kunduz, Afghanistan on 3 October 2015* (Report, United States Forces Command, 21 November 2015) <http://fpp.cc/wp-content/uploads/01.-AR-15-6-Inv-Rpt-Doctors-Without-Borders-3-Oct-15_CLEAR.pdf>.

**Part II – Liability in Civil and Criminal Law**

At this point, it is apposite to examine the concept of liability in both civil and criminal law, so that the requisite characteristics of that concept can be identified which are vulnerable to displacement by HMT. This displacement is likely to occur because of the bi-directionality of communication between the human and machine elements of the HMT, and subsequent reliance on that communication as a basis for taking action: a decision made by a machine or human, where one influences the other, has the potential to affect resulting liability.

Traditional Western legal systems attribute liability on a basis of 'the individual human person as the central unit of action and the appropriate object of blame'.[36] This idea of liability as blameworthiness, both factual and moral, informs how the civil and criminal look to achieve co-regulatory purposes by enforcing breaches of duties in ways that are generally complementary.[37] The criminal law attributes liability at a higher standard and burden of proof, acting in a more 'agent-focused' manner than the civil law, where mere negligence or breach of duty will suffice.[38] Though criminal liability might involve a similar assessment of compliance or non-compliance as civil law,[39] the criminal law is also a tool of social control designed to both punish those who have committed the offences as well as to virtue signal potential future offenders that such behaviour is anathematic to good policy.[40] The stigma of criminal convictions and incarceration also achieves a broader social effect than the necessity of remediating or repairing harm in the context of civil litigation.[41]

---

[35] As an example, the British Royal Air Force bombers colloquially known as "Dambusters" in the Second World War were not permitted to drop their bombs until the dams were in sight: Paul Brickhill, *The Dam Busters* (Macmillan, New York, 2017).

[36] Neha Jain, 'Autonomous weapon systems: new frameworks for individual responsibility', in Nehal Bhuta, Susanne Beck, Robin Geiβ, Hin-yan Liu, Claus Kreβ (Eds.) *Autonomous Weapons Systems: Law, Ethics, Policy* (Cambridge University Publishers, Cambridge, 2016), 303.

[37] Kenneth Simons, 'The Crime/Tort Distinction: Legal Doctrine and Normative Perspectives' (2007) 17 *Widener Law Journal*, 719.

[38] Peter Cane, 'Mens Rea in Tort Law' (2000) 20 *Oxford Journal of Legal Studies* 4, 533-556, 555.

[39] Andrew von Hirsch, Martin Wasik, 'Civil Disqualifications Attending Conviction: A Suggested Conceptual Framework' (1997) 56 *Cambridge Law Journal* 599, DOI:10.1017/S0008197300098597.

In this way, crimes outlaw particular activity and make it impermissible under every circumstance, whilst civil law prevents the breaches of rights and provides reparation of breaches – in economic terms, 'criminal law exclusively imposes *sanctions*, while [civil] law prices an *activity*'.[42] The two are not mutually exclusive – sometimes civil law can be used to punish, whilst criminal law can be used to remediate[43] – but the importance of criminal and civil systems remaining complementary and co-regulatory, but separate strands of law cannot be underestimated:[44]

> …it is a mistake to compare crime and tort. If three persons are incited by a fourth to break into a house and cause damage each will be guilty of a crime and will receive separate punishment. The inciter will be guilty of the criminal offence of inciting others to commit crime. The other three will be guilty of the crime of breaking in. If the damage [is] caused…then in a civil action the three who caused the damage will be jointly and severally liable… The inciter will also be jointly and severally liable for the damage if he procures the commission of the tort and is a joint tortfeasor.

Liability as blameworthiness

Liability as blameworthiness is thus a common cornerstone to both civil and criminal law, even if they are crafted and applied in different contexts.[45] In civil law, blameworthiness is usually established by applying common law principles such as taking reasonable care not to harm one's "neighbour" or person proximate to their conduct,[46] whereas for the criminal law it is the written Acts of some governing body such as Parliament or Congress that set out rules to be complied with.[47] In respect of making determinations of liability, the arbiter of law (the judge) and the arbiter of fact (often a judge but occasionally a jury) are called to offer an assessment of whether one party has broken a particular rule or breached a given duty.[48]

---

[40] Edoardo Greppi, 'The Evolution of Individual Criminal Responsibility Under International Law' (1999) 81 *International Review of the Red Cross* 531, 536-537; Rebecca Crootof, 'War Torts: Accountability for Autonomous Weapons' (2016) 164 *University of Pennsylvania Law Review* 6, 1347-1402

[41] Ibid.

[42] Robert Cooter, 'Prices and Sanctions' (1984) 84 *Columbia Law Review* 1523, 1523; see also Cane (n 38), 555.

[43] Cane (n 38).

[44] *CBS Songs Ltd v Amstrad Consumer Electronics Plc* [1988] UKHL 15; [1988] AC 1013.

[45] William J. Stuntz, 'Substance, Process, and the Civil-Criminal Line' (1996) 7 *Journal of Contemporary Legal Issues* 1, 19-24.

[46] *Donoghue v Stevenson* [1932] AC 562.

[47] Douglas Husak, *Overcriminalization: The Limits of the Criminal Law* (Oxford University Press, Oxford, 2008) 9-10.

Given the further social significance of a criminal finding of guilt (potentially involving the loss of an individual's liability through a custodial sentence) versus the pecuniary imposition of damages through establishing negligence, the standard of proof for criminal liability is objectively higher than in civil law. This concept is expressed in most legal systems as 'beyond reasonable doubt' as opposed to 'on the balance of probabilities',[49] and is expressed in somewhat equivocal terms in *Currie v Dempsey*:[50]

> In my opinion [the legal burden of proof] lies on a plaintiff, if the fact alleged (whether affirmative or negative in form) is an essential element in his cause of action, eg if its existence is a condition precedent to his right to maintain the action. The onus is on the defendant, if the allegation is not a denial of an essential ingredient in the cause of action, but is one which, if established, will constitute a good defence, that is, an "avoidance" of the claim which, prima facie, the plaintiff has.

Moral and physical blameworthiness is also imported into other terms used in the determination of liability. Upon assessment of a particular factual situation, questions may be asked around intent to engage in a particular act, which in turn invoke determinations of whether an action involves "strict" liability or whether liability is contingent upon finding a person held a particular state of mind – legally, the *mens rea* or "guilty mind".[51] It is only after exploring the complete factual situation that a person can be held responsible for some kind of illegal or wrongful act.[52]

This determination involves the importation of concepts of knowledge and intention to constitute moral blameworthiness, responsibility and punishment.[53] Put differently, the concept of intent provides for the ascription of blameworthiness, a reflection of the aphorism that 'an agent is responsible for all and only his intentional actions'.[54] Collectively, lawyers commonly talk of intent as both a mental state of intending some action, and intentionality of the action as motivated by that mental state.[55] Intentionality in criminal law has a very defined, and very precise, meaning and purpose: consisting of both the intention to engage in certain conduct and an intention to bring about a result because of that conduct (or knowledge that it will occur).[56] This is a deliberate choice: though "strict" liability exists in crime where no proof of intention is needed, it is usually reserved for minor or regulatory offences where the removal of proving intent is not considered procedurally unfair to the accused.[57] Equally, punishing only those offences that a person actually plans and then carries out severely constrains the legal system in regulating unlawful conduct.[58]

---

[48] Mike Redmayne, 'Standards of Proof in Civil Litigation' (1999) 62 *Modern Law Review* 2, 167-195.

[49] In Australia, this is discussed in the seminal case of *Briginshaw v Briginshaw* (1938) 60 CLR 336.

[50] (1967) 69 SR (NSW) 116, 539.

[51] Matthew R. Ginther, Francis X. Shen, Richard J. Bonnie, Morris B. Hoffman, Owen D. Jones, Rene Marois, Kenneth W. Simons, 'The Language of Mens Rea' (2014) 67 *Vanderbilt Law Review* 5, 1327-1372.

[52] *Vines v Djordjevitch* (1955) 91 CLR 512, 519.

[53] Bertram F. Malle, Sarah E. Nelson, 'Judging Mens Rea: The Tension between Folk Concepts and Legal Concepts of Intentionality' (2003) 21 *Behavioural Sciences and the Law,* 564.

[54] John L. Mackie, *Ethics: Inventing Right and Wrong* (Penguin UK, London, 1990), 208.

So whilst intentionality and intention may appear similar in both civil and criminal law, they are treated differently and can achieve different outcomes. Good motives cannot rescue or defend wrongful conduct, either in tort or crime. In *Caldwell*[59] an individual erected a wharf on public property and was charged with public nuisance. His defence – that the wharf was at the request of, and benefitted, the local community – was dismissed by the court because he had infringed a common right. On the other hand, a malign motive will taint any form of conduct, even if the conduct itself is morally acceptable. For example, a contract is a lawful arrangement between two parties and may be undertaken by any persons in society at large to regulate their dealings. However, a contract that is objectionable on public policy or legal grounds – such as a contract to commit murder – is void and unenforceable.[60]

Thus, the criminal law departs from civil law because the bare formulation of mental state and conduct grounds liability: there is no need to prove a particular effect or outcome. This explains the criminalisation of conduct even where both parties may consent (such as drug dealing or prostitution[61]),

---

[55] Malle and Nelson (n 53), 564.
[56] Issues of automatism and involuntariness are beyond the scope of this paper.
[57] David Prendergast, 'The Constitutionality of Strict Liability in Criminal Law' (2011) 33 *Dublin University Law Journal*, 285; cf. Federico Picinali, 'The denial of procedural safeguards in trials for regulatory offences: a justification' (2017) 11 *Criminal Law and Philosophy* 4, 681-703.
[58] Cane (n 38), 553.
[59] *Respublica v Caldwell* 1 U.S. (1 Dall.) 150 (Pa. Ct. of Oyer & Terminer 1785).
[60] *Commonwealth Bank of Australia Ltd v Amadio* (1983) 151 CLR 447.
[61] Barbara Sullivan, 'Rape, prostitution and consent' (2007) 40 *Australian & New Zealand Journal of Criminology* 2, 127-142.

where the offence never actually took place (such as attempting to commit a crime[62]) or where the offence was actually committed by someone else (inchoate crimes such as aiding or abetting, which are treated differently to contributory negligence[63]). Further, it is almost always the State – and not the infringed party – who brings proceedings for the commission of crimes.[64] Conduct might also be criminalised without reference to culpability if there was a serious social cost. In Blackstone's *Commentaries* he observed that the formulation of early "Crown" offences such as treason, carnal knowledge of the queen, piracy, serving a foreign monarch or harbouring a Catholic priest were punishable without any proof of intent.[65]

Conversely, the purpose of proving intention in civil law (especially torts) – as opposed to in criminal law, where intention may be a fundamental proof of the charge – may be unnecessary. Torts are almost always actioned by the aggrieved parties, and not the State, in order to receive remedies that place the aggrieved parties as near to their original position before the infringement.[66] Because the focus of tort liability is generally on the existence of a duty of care, a breach of that duty, and in most cases the suffering of harm, one cannot attempt a tort, plan one or conspire to cause one.[67] Intention is usually relevant to penalty, not liability; again, this is a deliberate choice. For the victim whose rights have been infringed, they might not necessarily care if an infringement was actuated by malice, recklessness or negligence. A search for intentionality may well be meaningless to compensating the harm the victim suffered.

That is not to say that intention in civil law is a dead or useless concept. Exemplary damages may be issued by the court in cases where the conduct was deliberately engaged in and 'of a sufficiently reprehensible kind'.[68] In this way, torts can "punish" intentional conduct in circumstances where an 'assertion of one's autonomy which, if it produces harmful consequences, may justify more onerous liability than negligence'.[69] Intention may also become more relevant where torts regulate activity that has a high social value but is inherently risky, such as transporting dangerous goods or manufacturing poisonous chemicals. In these contexts, it is apparent that the differences between negligence and malice are far more relevant to tortious conduct: in the words of Cane, 'when a harm-causing activity has high social value, a requirement of intention for tort liability helps to protect society's interest in the continuance of that activity'.[70]

---

[62] Ian D. Leader-Elliott, 'Framing preparatory inchoate offences in the Criminal Code: The identity crime debacle' (2011) 35 *Criminal Law Journal* 1, 80.
[63] Joachim Dietrich, 'The Liability of Accessories under Statute, in Equity, and in Criminal Law: Some Common Problems and (Perhaps) Common Solutions' (2010) 34 *Melbourne University Law Review* 1, 106.
[64] Ric Simmons, 'Private Criminal Justice' (2007) 42(1) *Wake Forest Law Review* 911.
[65] Guyora Binder, 'The Rhetoric of Motive and Intent' (2002) 6 *Buffalo Criminal Law Review* 1, 16-17.
[66] Scott Hershovitz, 'The Search for a Grand Unified Theory of Tort Law' (2017) 130 *Harvard Law Review* 3, 942-971; cf. Seth Davis, Christopher A. Whytock, 'State remedies for human rights' (2018) 98 *Boston University Law Review* 2, 397.
[67] Though a party may face contributory negligence for playing some part in its commission: Paul S. Davies, 'Accessory Liability for Assisting Torts' (2011) 70 *The Cambridge Law Journal* 2, 353-380; Paul S. Davies, Philip Sales, 'Intentional harm, accessories and conspiracies' (2018) 134 *The Law Quarterly Review*, 69-93.

Theoretical challenges to HMT liability

It is this focus on blameworthiness that will likely be disrupted by the appearance or adoption of HMT and its bi-directional communication between man and machine. In a legal system where the focus is on the punishment of unlawful conduct or the remediation of breaches of rights, any circumstance influencing the blameworthiness of an agent will have serious ramifications for attribution of liability: '[A]n agent can only be held responsible if they know the particular facts that surround their action, they are able to freely form a decision to act, and are able to select one of the suitable available alternative actions based on the facts of the given situation'.[71] Breaking apart this statement, we can consider three consecutive notions of attribution of liability and blameworthiness that are worth further exploration in the context of HMTs: **knowledge** of the facts, the existence of **suitable alternatives**, and the **freedom** to decide on one of them:[72]

- Knowledge: consider for a moment a HMT where the machine component of the team merely provides information or feedback to the human component, but the machine's programming suffers a catastrophic error and the feedback the human receives is completely nonsensical.

---

[68] *Lamb v Cotogno* (1987) 164 CLR 1.
[69] Cane (n 38), 553.
[70] Ibid.
[71] Matthias (n 29), 175.
[72] Giovanni Sartor, Andrea Omicini, 'The autonomy of technological systems and responsibilities for their use', in Nehal Bhuta, Susanne Beck, Robin Geiβ, Hin-yan Liu, Claus Kreβ (Eds.) *Autonomous Weapons Systems: Law, Ethics, Policy* (Cambridge University Publishers, Cambridge, 2016), 62.

Is a decision to engage in some form of conduct based on that erroneous information actionable if the conduct is subsequently proven to be unlawful?

- Suitable alternatives: Such an assessment is in part subjective and in part objective, considering what the individual thought suitable as well as the broader social context in which the act occurred.[73] Of course a lack of blameworthiness because of no suitable alternatives does not always exculpate the actor, as acts which are "the lesser of two evils" can still be an infringement on another's rights;[74]

- Freedom of choice: whether the human component of the HMT experienced a removal of freedom of choice will depend on the circumstances of the conduct and the context of the teaming operation. In circumstances of extreme emergency, or where the human and machine components are inextricably combined, there may be no way to divorce the human in any way that would render a valid freedom of choice. In others, the factual circumstances are highly relevant to blameworthiness: a military HMT operation in a combat zone may result in far less freedom of choice than an administrative HMT operation within an office.

By examining and weighing all three concepts we suggest it is possible to assess the degree to which a HMT might be liable for acts undertaken, and appropriately adjusting for the artificial component's effect on human decision making. The actions of within a HMT, assessed partly on actions by a machine and partly on actions by a human, will become enmeshed to varying extents and may lead to overlapping spheres of liability for blameworthiness. Given the nature of HMTs, it is perhaps easiest to 'conceive of their actions as creating a web of overlapping chains of responsibility, both criminal and civil in nature'.[75] As a concept, this idea already appears in the literature in the context of attributing liability to autonomous systems more generally:[76]

> The *mens rea* of the direct perpetrator therefore must be judged in terms of the secondary party's mental state, and will require intent or knowledge. This can also apply to AWS, as their code gives them the ability to perform some decision-making capabilities, and therefore be able to comprehend certain elements of their actions. However, ultimately, their actions are limited by a human agent, who sets parameters for how they are able to act. Therefore, responsibility can be shared by both the AWS and another human counterpart who is involved in its behaviours and actions.

---

[73] Randolphe Clarke, 'Moral responsibility, guilt, and retributivism' (2016) 20 *The Journal of Ethics* 1, 121-137, 124.

[74] James Goudkamp, 'The Spurious Relationship Between Moral Blameworthiness and Liability for Negligence' (2004) 28 *Melbourne University Law Review* 2, 343.

[75] Jain (n 36), 304.

[76] Jain (n 36), 310.

Pragmatic challenges to HMT liability

There are also some unresolved difficulties in the application of liability and blameworthiness to HMT more generally. The first is identifying which actor within a HMT, whether the machine or human actor, is the one "making" a decision when the tasks being completed are not repetitive or deterministic.[77] Consider the theoretical effects explored above: what if a human is presented with a tactical scenario in which there are no alternative options which the human considers acceptable. If the human takes what is considered to be the only "reasonable" option, are they really making a decision? Or has the decision already been made by the machine – perhaps inadvertently – by presenting the information in a way only one option was possible?

The second challenge, particularly in the military and armed forces context, is the effect of HMTs on inquisitorial process (such as criminal investigation or civil discovery). Often these processes involve determining both a factual substrate of the conduct, but also an assessment of liability. Unfortunately, HMT presents two distinct barriers to these processes. Firstly, the fact that much of the technology, automation and/or software underpinning HMTs is likely to be protected by trade secrets or military secrecy;[78] and secondly, the opacity of AI/automation programs in HMT means that even where such the code of such programs can be exposed, the apparent nature of decision-making by that code is not readily discernible in a manner understandable by jurors or judges.[79]

The third is the differing legal treatment of various mental defences within and across jurisdictions. It is not within the scope of this article to consider the various natures of impairment, automatism or insanity defences (however they might be labelled); instead, it is to note that the varying degrees, scope and application of these defences will lead to entirely varied treatments of HMTs in circumstances where judges are called to assess the "voluntariness" of actions to assign blameworthiness.[80] This is especially the case where many of the mental defences often involve some level of "impairment" to functioning – does the human in a HMT really become "impaired" because of the inclusion of a machine component?[81]

---

[77] Shagun Jhaver, Iris Birman, Eric Gilbert, Amy Bruckman, 'Human-machine collaboration for content regulation: The case of reddit automoderator' (2019) 26(5) *ACM Transactions on Computer-Human Interaction* 1, 8; Vincent Boulanin, Neil Davison, Netta Goussac, Moa Peldán Carlsson, *Limits on Autonomy in Weapon Systems: Identifying Practical Elements of Human Control* (Report for the Stockholm International Peace Research Institute, 2020).

[78] Gabriele Spina Ali, Ronald Yu, 'Artificial Intelligence between Transparency and Secrecy: From the EC Whitepaper to the AIA and Beyond' (2021) 12(3) *European Journal of Law and Technology* 1, 6-8.

[79] Ashley Deeks, 'The judicial demand for explainable artificial intelligence' (2019) 119(7) *Columbia Law Review* 1829.

The last challenge for regulating HMTs in a pragmatic sense is determining a remedy that adequately reflects the blameworthiness of the conduct. Most Western legal systems have evolved from the perspective that irrespective of the legal entity a claim is brought against, there is nevertheless a 'human who decides whether or not to comply'.[82] For example, civil and criminal actions are often brought against companies as legal entities, but where the actions of those companies are often a sheeted home to individuals within them.[83] Where a machine can be attributed with blameworthiness, there comes the question of how to achieve a penalty or restitution in a manner that is relevant to the machine. Alternately, there is a question of how to apply a remedy to a human who may have had no conscious control of or over the actions they are now alleged to have engaged in.[84]

In summary, how might these various principles of theoretical and pragmatic challenges in the context of military HMTs be accounted for? Whatever the intended scheme of regulation is proposed, we consider that it must be capable of addressing the difficulties of applying blameworthiness in the context of HMT operations generally, but also the military and armed forces context more specifically. We consider the best and most efficient approach to involve modifying an existing regulatory scheme to apply to the future use of military HMT operations.

---

[80] Peter Cane, 'Fleeting Mental States' (2002) 59(2) *Cambridge Law Journal* 273.

[81] These defences are generally 'exculpatory doctrines' to apply in only 'the most extreme cases': Steven Yannoulidis, 'Excusing Fleeting Mental States: Provocation, Involuntariness and Normative Practice' (2005) 12(1) *Psychiatry, Psychology and Law* 23, 27.

[82] Lawrence Lessig, 'The Zones of Cyberspace' (1996) 48 *Stanford Law Review* 1403, 1408.

[83] Michael Nietsch, 'Corporate illegal conduct and directors' liability: an approach to personal accountability for violations of corporate legal compliance' (2018) 18(1) *Journal of Corporate Law Studies* 151.

[84] David Watson, 'The rhetoric and reality of anthropomorphism in artificial intelligence' (2019) 29(3) *Minds and Machines*, 417.

## Part III – A Proposed Framework for Liability in Human-Machine Teams

In this final Part, we propose the leveraging of a concept that has already been explored in the literature – chains of responsibility or "COR" – as a mechanism for attributing liability in HMT. Originating in the logistics and supply chain industry, COR applies a proactive model of compliance to prevent road and freight accidents. COR legislation for heavy vehicles is already a feature of the legal landscape in Australia.[85]

It is with this framework in mind that we present a COR model for the HMT context in Table 1. Along the vertical axes, Table 1 charts the lifecycle of a HMT from conception and design, through manufacture and testing, to procurement and deployment (both domestic and foreign). At each stage of that lifecycle, those involved with HMT will carry responsibilities explored horizontally. These responsibilities are non-exhaustive, intended to provide a high-level example of the types of activities at each stage which are relevant in determining potential legal culpability from HMT use. Each of them has been derived from the theoretical and pragmatic challenges to HMT liability in Part II and are intended to be read from perspective of "reasonable foreseeability". For example, legal and ethical advice in the conduct of military operations is best sought well before the first shot is fired, when advising how a conflict may be fought[86] - thus legal and ethical advice should be incorporated into the very design of the HMT.[87] To discharge their responsibilities at manufacture, those producing HMT interfaces and software should have in place rigorous testing regimes capable of detecting flaws and errors to a low tolerance (noting that the systems will ultimately be used in a warfighting capability[88]). Those involved in the manufacture of subcomponents will also need to meet these rigorous standards and be informed by the principal manufacturer of the potential risks.[89]

---

[85] Heavy Vehicle National Law (Queensland); as applied by the *Heavy Vehicle National Law Act 2012* (Qld), s 4; *Heavy Vehicle National Law Act 2013* (ACT), s 7; *Heavy Vehicle (Adoption of National Law) 2013* (NSW), s 4; *Heavy Vehicle National Law (South Australia) Act 2013* (SA), s 4; *Heavy Vehicle National Law (Tasmania) Act 2013* (TAS), s 4; *Heavy Vehicle National Law Application Act 2013* (Vic), s 4

[86] Marcus Schulzke, 'Autonomous Weapons and Distributed Responsibility' (2013) 26(2) *Philosophy & Technology* 203, 209.

[87] Heather M. Roff, 'The Strategic Robot Problem: Lethal Autonomous Weapons in War' (2014) 13(3) *Journal of Military Ethics* 211.

[88] Jai Galliott, 'The Soldier's Tolerance for Autonomous Systems' (2018) 9(1) *Journal of Behavioral Robotics* 124.

[89] Brendan Walker-Munro, 'Exploring manufacturer strict liability as regulation for autonomous military systems' (2022) 27(1) *Torts Law Journal* 182

**Chain of Responsibility for HMT**

| | | | |
|---|---|---|---|
| **Design Phase** | Obtain legal and ethical advice regarding design. | Critically examine whether system functions being automated are appropriate for the intended purpose. | Build compliance with international and domestic law and regulation into planning and design phases. |
| **Manufacturing** | Advise subcontractors of any potential liability for including their components. | Testing of components must be at or above industry standard and meet regulatory requirements where applicable. | Adverse incidents must be fully investigated to eliminate possibility of safety defects. |
| **Testing** | Testing must be frequent and robust across all proposed operating environments. | Any potential safety concerns that cannot be remediated must be included as mandatory warnings | Safety of the platform must be "such as persons generally are entitled to expect". |
| **Contract Negotiation** | Include, as contractual terms, the intended scope and operating environment/s of the platform/s. | Disclose all possible safety issues with the platform. | Maintenance and upkeep of platforms will continue to incur potential liability. |
| **In-Service** | Testing must be compatible with intended operation – do not rely on manufacturer testing. | "Handover" to Defence does not end the possibility for liability for safety defects. | Operators must not operate systems until deemed competent. |
| **Domestic Deployment** | Platform must be capable of operating consistent with domestic and international laws. | Human decision-maker must be capable of intervening at any time. | Systems must be operated in accordance with manufacturer's instructions and training at all times. |
| **International Deployment** | Be aware that Australian and foreign domestic law will apply to the use of the platform. | Human decision-maker must be capable of intervening at any time. | Consider information and technology security in overseas environments. |

**Table 1: Proposed nature of COR applying to HMT**

The concept of COR is relatively simplistic: it imposes a primary duty on every person interacting with a given shipment to ensure the safety of their activities so far as is reasonably practicable.[90] In the event of an accident or incident, an investigation is conducted that examines the entire logistic chain to determine where the duty was breached, and by which agent. Breaches of that duty of care may result in the commencement either of civil action (involving pecuniary penalties) or criminal offences (involving potential for penal sentences in severe cases). There exists a legitimate question as to how COR might actually address any of the theoretical or pragmatic challenges that HMT proposes to existing civil and criminal liability approaches. It is therefore the focus of this Part of the article to demonstrate how COR could apply in the context of military HMTs.

Applying COR to theoretical challenges to liability

It should be recalled that we examined three consecutive notions of attribution of liability and blameworthiness applying to HMTs: **knowledge** of the facts, the existence of **suitable alternatives**, and the **freedom** to decide on one of them.[91] How should COR apply to these three notions of liability?

Firstly, COR examines the nature of actions taken and decisions made up to and inclusive of the decision to engage in the impugned conduct. In such an *ex ante* examination, it is not just the blameworthiness of the ultimate decision that is determinative, but each of the "steps" which led up to its final execution. Consider our earlier example of a military HMT where the machine programming fails and the human is presented with nonsensical information. In this case, the application of COR's "reasonably practicable" assessment of safety might determine that the nature of the manufacturer's pre-deployment testing was insufficient, and that this was the blameworthy failure. Alternately, it might be a repairer who inserted a faulty component who bears the blame for the incident at hand. This concept of extended liability is certainly not unknown to the literature, and reinforces the idea that delictual responsibility is not a pie – '[a]ll involved can theoretically take all the responsibility for the harm caused, and if unjustified, punished'.[92]

---

[90] Amanda Beesley, *Improving safety and compliance, and simplifying enforcement–recent reforms to Australia's heavy vehicle chain of responsibility laws* (Conference paper, HVTT14: International Symposium on Heavy Vehicle Transport Technology, Rotorua, New Zealand, 5 October 2016); Geoff Farnsworth, Jarrad McCarthy, 'Heavy Vehicle National Law reform: New approach to chain of responsibility liability' (2016) 68 *Governance Directions* 1, 41-43; Wonmongo Lacina Soro, *Towards an understanding of financial influences on heavy vehicle safety outcomes* (PhD thesis, Queensland University of Technology, 2020).

[91] Sartor and Omicini (n 72).

Secondly, whilst COR applies equally across the nature of military and non-military actors, it has the flexibility to consider the unique challenges of military service. The application of COR is based on precautions which are "reasonably practicable" by reference to the controls available, the suitability of those controls, and the cost of controls proportional to the risk posed. Consistent with other approaches to applying liability to emerging military technologies, it is easier to control risks from a manufacturer's office or a designer's factory, places which are far removed from operations against the enemy in a foreign war zone.[93]

Thirdly, COR sidesteps many of the theoretical issues to liability attribution which might occur in the context of military HMT. Again, the curial search in COR is for "reasonable practicability", not necessarily strict blameworthiness. In circumstances where a military force has not properly trained a human for HMT operations but could have easily done so with the resources and time it had available, then underlying fault questions does not arise. The military force bears responsibility under COR and may be prosecuted or litigated accordingly. Military forces are already assessed for this level of compliance under most Western work health and safety systems, suggesting that the level of adaptation required to adopt COR is unlikely to be onerous or disruptive to military operations.[94]

Fourthly, by adopting a less prescriptive system for attribution of blameworthiness, there is potential to avoid the 'injustice' of liability being applied to persons not having sufficient control of machines or in circumstances where the machine has malfunctioned.[95] At the same time, COR renders irrelevant the need for militaries to scrutinise the role of humans in a decision cycle[96] – the focus of inquiry in situations of failure is on the reasonableness of safeguards enacted to protect against harms, not the actions of the individual HMT.

---

[92] Ross W. Bellaby, 'Can AI Weapons Make Ethical Decisions?' (2021) 40(2) *Criminal Justice Ethics* 86, 96.
[93] *Smith v Ministry of Defence* [2013] UKSC 41; Walker-Munro (n 89), 182.
[94] Nick Turner, Sarah J. Tennant. '"As far as is reasonably practicable": Socially constructing risk, safety, and accidents in military operations' (2010) 91(1) *Journal of Business Ethics* 21; John A. Casciotti, 'Fundamentals of military health law: governance at the crossroads of health care and military functions' (2016) 75(1) *Air Force Law Review* 201.
[95] Matthias (n 29), 183.
[96] Ryan (n 30).

Finally, COR is equally adaptable to the vast array of field environments which modern militaries are prepared to operate in. The reasonableness of safety precautions to avoid specific harms recognises that 'a safety measure that would be enough in one situation might be completely inadequate in another and excessive in a third set of circumstances'.[97] Therefore, what may be deemed acceptable to limit HMT risk in a training setting might be inadequate for foreign operations but excessive in joint or allied exercises – dependent entirely on the circumstances of the HMT deployment and the likelihood and magnitude of the risk being guarded against.

Applying COR to pragmatic challenges to HMT liability

In the same vein, COR has the potential to drastically limit or eliminate the pragmatic risks to the attribution of HMT liability. The application of COR to HMT operations, especially military operations, recognises the unique factual circumstances in each deployment of HMT and seeks to impose a sliding duty of reasonableness to determine whether liability should apply and to what degree.

Firstly, the idea of needing to determine which aspect of an HMT – human or machine – "made" a decision for the purposes of any liable conduct is irrelevant. The focus of COR is not strictly limited to the liable conduct in question, but on all the antecedent decisions and circumstances along the chain leading to that conduct. Where a programming error which presents inaccurate or misleading data in an HMT is the causative agent, liability is still attributable to the human for not verifying the information using another technique or system. The fighter pilot who bombs a target without visually verifying and satisfying themselves of a target's validity (and instead relying on the automated or autonomous system) stands to carry some of the punishment or rectification for that fault.

Secondly, COR considers but does not rely on the mental element of each of the individual actors along the chain. Each individual actor has the same duty ("to limit risk to extent reasonably practicable") but different capacities and methods of discharging that duty, in the same way as modern work health and safety laws operate in the military context.[98] Defences of automatism, involuntariness or diminished capacity are relevant only to the extent that the actor can discharge the duty, not the existence of the duty. It further recognises that the liability of one individual of the chain may be contingent, or rely upon, the liability of others – the failure of a soldier relying on faulty data is influenced by and partly reliant upon the failure of a manufacturer to properly test the machine components.

---

[97] National Heavy Vehicle Regulator, 'Primary duty definitions', *Safety, Accreditation and Compliance* (website, 2022) <https://www.nhvr.gov.au/safety-accreditation-compliance/chain-of-responsibility/the-primary-duty/primary-duty-definitions>.

Thirdly, COR has the potential to move with new developments in technology, including the ability for machines to achieve "artificial general intelligence".[99] At such a point, machine components in HMTs may well be treated as moral actors with their own level of agency, at which point they become another link in the COR and their potential for blameworthiness becomes examinable. If a machine can achieve general intelligence in a manner that can be attributed moral agency, there is no reason why its actions could not be examined through the lens of reasonable practicability for preventing harm.

Of course, COR has limited ability to counter the challenges of trade secrets, secrecy or explainable AI; indeed, it potentially introduces a challenge in the form of the broadly conceptualised "state of the art" defence. This defence obviates responsibility in COR and similar regimes for defects which could not be detected by reasonably practicable testing available at the time of manufacturing or programming.[100] However, many Western legal systems have grappled with like concepts for decades. In most cases they involve interpretation and application of rules of procedure which are dealt with at the level of individual courts or tribunals (including their military equivalents). Vesting arbiters of fact with appropriate powers of inquiry, coupled with broader education of the legal fraternity in concepts like AI, are already subject to calls for action across multiple domains.[101]

---

[98] Turner and Tennant (n 99).

[99] Ben Goertzel, 'Artificial general intelligence: concept, state of the art, and future prospects' (2014) 5(1) *Journal of Artificial General Intelligence* 1; Pei Wang, 'On defining artificial intelligence' (2019) 10(2) *Journal of Artificial General Intelligence* 1.

[100] Mabel Tsui, 'The state of the art defence: defining the Australian experience in the context of pharmaceuticals' (2013) 13(1) *QUT Law Review* 132.

[101] Deeks (n 79).

## Conclusion

There can be little doubt that the integration of human and machine elements offer significant benefits to the armed forces, yet there has not been sufficient consideration of how we might regulate those integrations. Given the significance of decisions made in the context of military operations, which might involve the deaths of hundreds or thousands of people, we cannot leave the regulation of such events to mere chance and ambiguity. Nor does there appear to be much benefit in merely outlawing the pursuit of HMT applications, driving the research underground and delegitimizing a purposeful line of human research.

Instead, what is required is a nuanced and purposeful regulatory regime which considers the reasoning for attribution of responsibility, whilst also providing appropriate mechanisms for restitution and punishment. This is much for the benefit of our armed forces as for the protection of the rules-based global order: military officers and personnel need to know the legal limits of their conduct, what can be done in war and peace time, and what consequences might attach when they step outside those boundaries.

There is more yet to be done. The exact parameters of technologies designed to constitute HMT and how they are defined in law will need more comprehensive examination than was possible in this article. The definitions will need to be expansive enough to capture those technologies at the forefront of military and civilian research, but also those yet to be contemplated. Alternately, new legal definitions for those technologies will need to be included in their own regulatory regime to eliminate grey areas and ambiguity. Just like our treatment AI, we need to ensure that the definition is clear, unambiguous and is not leading to inaccurate or oversimplified definitions of the technology.[102]

The work on regulating HMT also will not end with the possible introduction of a COR regime. There will no doubt be developments in warfighting technology which escape even the most carefully drafted working definitions of HMTs. Systemic difficulties which cannot be resolved at the procedural level of courts and tribunals will inevitably arise. Future avenues of research might look at how COR regimes could be tailored to specific military operations, or how military COR might be adapted to civilian environments. At the same time, broader calls for "explainable AI", rules of evidence for AI and judicial education need to be heeded to ensure that any COR regime enacted by an armed force is capable of being properly and justifiably dealt with.[103]

---

[102] Wang (n 99).

---

[103] Katherine Quezada-Tavárez, Plixavra Vogiatzoglou, Sofie Royer, 'Legal challenges in bringing AI evidence to the criminal courtroom' (2021) 12(4) *New Journal of European Criminal Law* 531.